\documentclass[twocolumn]{jpsj3}
\usepackage{txfonts}
\usepackage{bm}
\usepackage{url}
\usepackage[dvipdfmx]{graphicx}
\usepackage{color}

\title{Performance Evaluation of Associative Watermarking Using Statistical Neurodynamics}

\author{Ryoto Kanegae$^1$
  and Masaki Kawamura$^1$\thanks{kawamura@sci.yamaguchi-u.ac.jp}}
\inst{$^1$Graduate School of Sciences and Technology for Innovation,
Yamaguchi University, Yamaguchi, Yamaguchi 753-8512, Japan}

\abst{
We theoretically evaluated the performance of our proposed associative watermarking method in which the watermark is not embedded directly into the image. We previously proposed a watermarking method that extends the zero-watermarking model by applying associative memory models. In this model, the hetero-associative memory model is introduced to the mapping process between image features and watermarks, and the auto-associative memory model is applied to correct watermark errors. We herein show that the associative watermarking model outperforms the zero-watermarking model through computer simulations using actual images. In this paper, we describe how we derive the macroscopic state equation for the associative watermarking model using the Okada theory. The theoretical results obtained by the fourth-order theory were in good agreement with those obtained by computer simulations. Furthermore, the performance of the associative watermarking model was evaluated using the bit error rate of the watermark, both theoretically and using computer simulations.
}
\kword{associative memory, statistical neurodynamics, retrieval process, storage capacity, watermarking}
\begin{document}
\maketitle

\section{Introduction}
Illegal use of digital content in social media is a serious problem. In particular, unauthorized use and tampering with images and videos have been frequently detected. The digital watermarking method~\cite{petitcolas1999information,mohanty1999digital,mohanarathinam2020digital} is an effective method to solve such problems. In this kind of method, information such as copyrights and licenses is imperceptibly embedded in digital content. This prevents unauthorized use of the content and identifies the authorized owner. The information embedded in the content is called a watermark. Embedding methods include pixel substitution~\cite{celik2005lossless}, which adds the watermark to a pixel, and embedding in the frequency domain obtained by discrete cosine transformations (DCT)~\cite{cox1996secure,cox1997secure} and other methods~\cite{luo2011robust}. These embedding methods place the watermark directly into the content, causing distortion in the image.

The zero-watermarking method~\cite{dong2011robust,rani2015zero,kim2015crt} was proposed as a method of undistorted image watermarking. In this method, instead of embedding a watermark in an image, a secret key is generated from the watermark and the unique features extracted from the image. The validity of the image can be determined by retrieving the watermark from the features and the stored secret key. Because this method never embeds a watermark, the image is never distorted. However, the watermark cannot be retrieved correctly if the original image is attacked because there is no error correction capability. Furthermore, if the zero-watermarking method were applied to a large number of images, a large number of secret keys would be generated, requiring that the pair between the secret keys and features be managed.

The hetero-associative memory model (HMM)~\cite{kawamura1999dynamics} can store pairs of key patterns and associative patterns, and the corresponding associative patterns can be recalled if a key pattern is input. Therefore, the HMM can be applied to achieve the zero-watermarking method by considering the features extracted from the images as key patterns and by considering the watermarks as associative patterns. Applying the HMM has the following advantages. First, the sizes of the features and watermark can be made different. Second, if the features have errors, the HMM can correct the errors in the watermark. Third, pair management is no longer necessary because multiple pairs can be stored in the HMM. The auto-associative memory model (AMM)~\cite{hopfield1982neural,amari1988statistical,okada1995hierarchy} stores multiple patterns, and the most similar pattern among the stored ones is recalled when a pattern is input. Therefore, watermark errors can be corrected by storing watermarks in the AMM.

We proposed an associative watermarking method (AWM) based on the HMM and AMM to enable a zero-watermarking method~\cite{kanegae2022proposal}. The AWM is composed of a HMM in the first layer and an AMM in the second layer, resulting in the same network structure as the human associative processor (HASP)~\cite{hirai1983model,kawamura1997storage}. Owing to the error correction capability of the associative memory models, we showed using computer simulations that the watermark is retrieved without errors even when the image is attacked. In this paper, we present an evaluation of the error correction capability of the AWM through theory and computer simulations. Because AWM consists of the associative memory models, its recall process of the watermark and memory capacity can be obtained by using statistical neurodynamics, in particular, Okada theory~\cite{amari1988statistical,okada1995hierarchy}. The macroscopic state equations for the HASP have been derived~\cite{kawamura1999dynamics}.
Since the structure of the proposed model is the same as that of HASP~\cite{kawamura1999dynamics}, its macroscopic state equations were derived. The recall process for one-to-many associations was analyzed in the previous study~\cite{kawamura1999dynamics}. In this paper, we present new results to characterize it as the AWM.
We evaluated the bit error rate of the watermark by using the equations.

The paper is organized as follows. Section 2 explains the zero-watermarking method. Section 3 discusses associative memory models. Section 4 describes the associative watermarking method. Section 5 describes computer simulations that were conducted to evaluate the theory. Section 6 concludes the paper.

\section{Zero-watermarking method}
We explain the mechanism of the zero-watermarking method using DCT coefficients as features~\cite{dong2011robust}. Let $\bm{\xi}=\left(\xi_1,\xi_2,\cdots,\xi_K\right)^\top$ be a $K$-bit watermark, where each element of the watermark $\xi_i$ takes the value $\pm1$.

\begin{figure}[t] \centering
  \includegraphics[width=80mm]{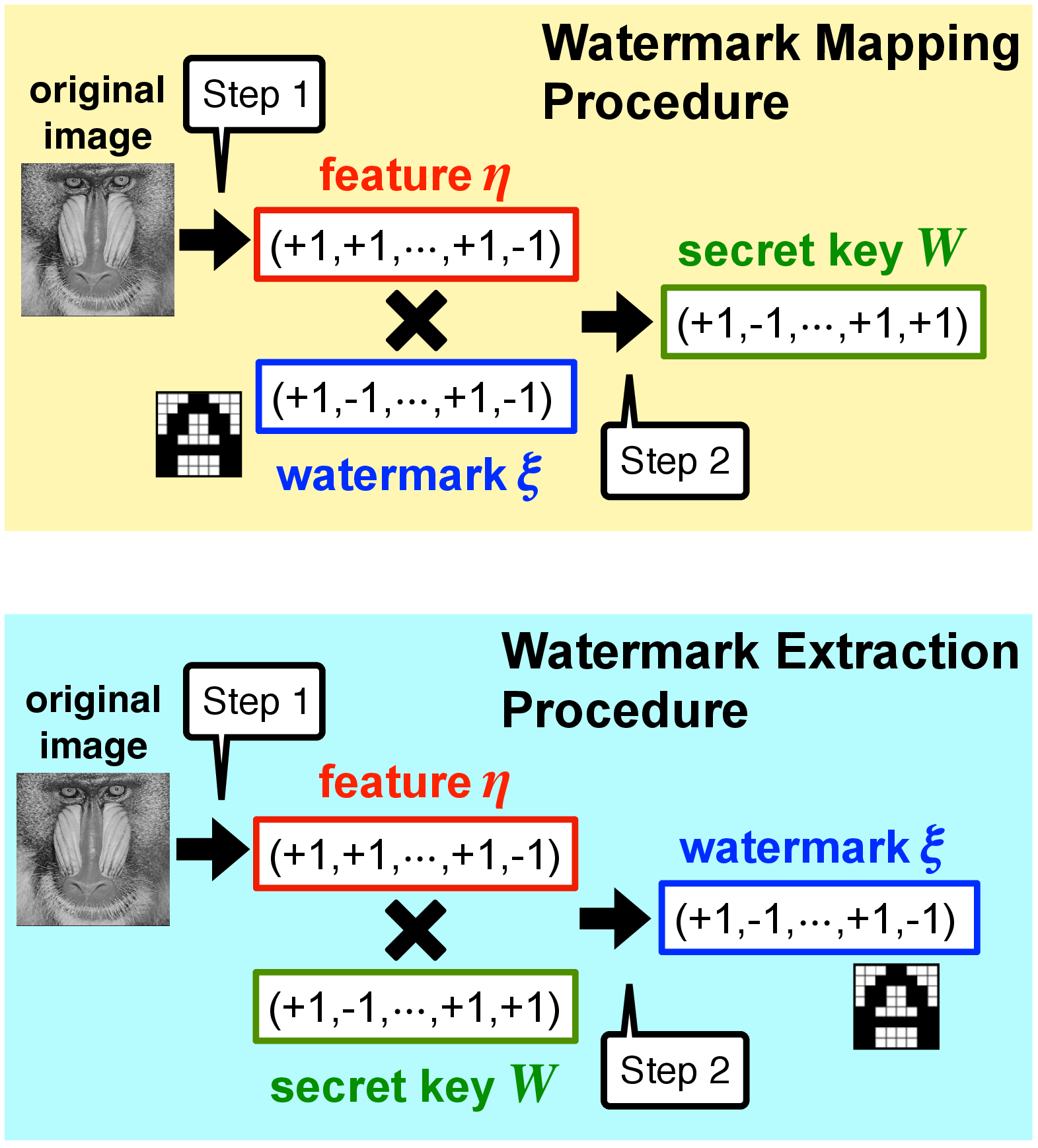}
  \caption{(Color online) Mechanism of zero-watermarking method.} \label{fig:Zero Watermarking}
\end{figure}

\subsection{Mapping procedure from the features and watermark to the secret key}
As shown in the upper part of Figure~\ref{fig:Zero Watermarking}, mapping from features and a watermark to a secret key involves two steps.

\subsubsection*{\textbf{Step 1}: Extract features from the original image.}
Two-dimensional DCT is performed on the original image. Low-frequency components are chosen as features because the changes in the DCT coefficients are small when the image is degraded. $K$ low-frequency components $\bm{d}=\left(d_1,d_2,\cdots,d_K\right)^\top$ are extracted from the DCT coefficients, excluding the DC component. The feature $\bm{\eta}=\left(\eta_1,\eta_2,\cdots,\eta_K\right)^\top$ is obtained by binarizing these extracted DCT coefficients $\bm{d}$, that is, given by
\begin{eqnarray} \label{sgn}
  \eta_i &=& \mathrm{sgn}\left(d_i\right),\; i=1,2,\cdots,K,
\end{eqnarray}
where the sign function $\mathrm{sgn}\left( \cdot \right)$ is defined as
\begin{eqnarray}
  \mathrm{sgn}(x) &=& \begin{cases}
    {+1,\ x \geq 0}\\
    {-1,\ x < 0}
    \end{cases}.
\end{eqnarray}

\subsubsection*{\textbf{Step 2}: Generate a secret key.}
The secret key $\bm{W}$ is generated by the Hadamard product of the feature $\bm{\eta}$ and the watermark $\bm{\xi}$. That is, it is calculated in Step 1. That is,
\begin{eqnarray}
  W_i &=& \eta_i \xi_i,\; i=1,2,\cdots,K.
\end{eqnarray}
Note that the bit length of the feature, $K$, must be the same as the bit length of the watermark. Finally, the generated secret key is stored.

\subsection{Watermark extraction procedure}
As shown in the lower part of Figure~\ref{fig:Zero Watermarking}, the procedure for obtaining a watermark from a feature and a secret key consists of two steps.
\subsubsection*{\textbf{Step 1}: Extract features from the original image.}
This step is identical to step 1 in the previous subsection.

\subsubsection*{\textbf{Step 2}: Extract a watermark.}
A watermark $\bm{\xi}$ can be obtained by the Hadamard product of a feature $\bm{\eta}$ and a secret key $\bm{W}$ . That is, it is calculated by
\begin{eqnarray}
  \xi_i &=& W_i \eta_i,\; i=1,2,\cdots,K.
\end{eqnarray}

Thus, the zero-watermarking method can map a watermark to a feature derived from the original image by generating a secret key. Because no watermark is embedded in the image, i.e., the image itself is not modified, no degradation in image quality occurs. However, the zero-watermarking method cannot extract the watermark if the original image has been degraded. Therefore, if compression, noise addition, filtering, or other attacks are applied to the image, the extracted feature might not coincide with the original, and errors might occur in the watermark. In other words, the method lacks robustness against attacks. In addition, because the Hadamard product is used to generate the secret key, there is a restriction that the bit length of the features and the watermark must be the same. Furthermore, as the number of images increases, the number of secret keys also increases. Therefore, it becomes difficult to manage the pairing between the features and the secret keys.

\subsection{Amount of information required to store}
Let us calculate the amount of information required to store the secret keys, $C_z$, in the zero-watermarking method. The $K$-bit secret key must be stored.  If $P$ secret keys are to be stored, $PK$ bits of information must be stored. In addition, information about the mapping of images to secret keys must be stored. Note that the need for this mapping information is not considered in conventional zero-watermarking methods~\cite{dong2011robust,rani2015zero,kim2015crt}. We assume that an index of $L=\log_2P$ bits is appended to both the image and the secret key to manage them. The index allows us to find the secret key corresponding to the image. This management information requires $2LP$ bits. Thus, the total information to be stored is $C_z=PK+2LP=P(K+2\log_2P)$ bits. However, if the use of the index is allowed, the watermark can be indexed directly. It can be easily mapped to the image. As a result, it is no longer necessary to implement the zero-watermarking method.  The zero-watermarking method does not support the management of multiple images.

\section{Associative memory models}
There are two types of associative memory models: the hetero-associative memory model~\cite{kawamura1999dynamics} and the auto-associative memory model~\cite{hopfield1982neural,amari1988statistical,okada1995hierarchy}.

\subsection{Hetero-associative memory model}
The HMM stores mappings from key patterns to associative patterns. Therefore, when a key pattern is given to the input layer, the corresponding associative pattern is recalled in the output layer. Let $\bm{\eta}^{\mu}$ be the $\mu$-th key pattern and let $\bm{\xi}^{\mu},\mu=1,2,\cdots,P$ be the $\mu$-th associative pattern, where $P$ is the number of patterns. The bit length of the key pattern is $K$ bits, and that of the associative pattern is $N$ bits. That is, the key pattern and the associative pattern are represented by $\bm{\eta}^\mu=\left( \eta^\mu_1,\eta^\mu_2,\cdots,\eta^\mu_K \right)^\top$ and $\bm{\xi}^\mu=\left( \xi^\mu_1,\xi^\mu_2,\cdots,\xi^\mu_N \right)^\top$, respectively. We assume that each component of the key and associative pattern takes the value $\pm1$ with equal probability and is given by
\begin{eqnarray}
  \label{eq:Peta}
  \mathrm{Prob}\left[\eta^\mu_i = \pm 1 \right] &=& \frac{1}{2},\\
  \label{eq:Pxi}
  \mathrm{Prob}\left[\xi^\mu_i = \pm 1 \right] &=& \frac{1}{2}.
\end{eqnarray}
The synaptic weight $W^h_{ik}$ of the HMM is given by
\begin{eqnarray} \label{W^h}
  W_{ik}^h &=& \frac{1}{N} \sum_{\mu=1}^P \xi_i^\mu \eta_k^\mu.
\end{eqnarray}
As an input $\bm{y}=\left(y_1,y_2,\cdots,y_K\right)^\top$ is given to the HMM, the output of the HMM, $\bm{x}^0=\left( x_1^0,x_2^0,\cdots,x_N^0 \right)^\top$, is given by
\begin{eqnarray} \label{hetero_output}
  x^0_i &=& \mathrm{sgn}\left(h_i\right),
\end{eqnarray}
where $h_i$ is the internal state of the neuron and is defined as
\begin{eqnarray} \label{internal state_hetero}
  h_i &=& \sum_{k=1}^K W^h_{ik} y_k.
\end{eqnarray}
Here, we define the similarity between the input $\bm{y}$ and the $\nu$-th key pattern $\bm{\eta}^\nu$ as overlap given by
\begin{eqnarray} \label{overlap_feature}
  m^\nu_* &=& \frac{1}{K} \sum_{k=1}^K \eta_k^\nu y_k.
\end{eqnarray}
The internal state given by (\ref{internal state_hetero}) can be expressed as
\begin{eqnarray} \label{interference term_hetero}
  h_i &=& \gamma m^\nu_* \xi^\nu_i + z_i^*,
\end{eqnarray}
where $\gamma=\frac{K}{N}$ and $z_i^*$ are the crosstalk noise term, given by
\begin{eqnarray} \label{crosstalknoise_hetero}
  z_i^* &=& \frac{1}{N} \sum_{k=1}^K \sum_{\mu \neq \nu}^{P} \xi^\mu_i \eta^\mu_k y_k.
\end{eqnarray}
The crosstalk noise $z_i^*$ is subject to a Gaussian distribution with mean $0$ and variance $\sigma^2_*$ when the large system limit $N\to\infty,K\to\infty,P\to\infty$ hold and both the loading rate $\alpha=\frac{P}{N}$ and $\gamma$ are finite~\cite{kawamura1999dynamics}. The variance $\sigma^2_*$ given by
\begin{eqnarray}
  \sigma^2_* &=& E\left[\left(z_i^*\right)^2\right] = \alpha\gamma. 
\end{eqnarray}
The overlap $m^\mu_0$ between the output $\bm{x}^0$ of the HMM and the $\mu$-th associative pattern $\bm{\xi}^\mu$ is defined by
\begin{eqnarray} \label{overlap_hetero}
  m_0^\mu &=& \frac{1}{N} \sum_{i=1}^N \xi_i^\mu x^0_i.
\end{eqnarray}
In the large system limit, the overlap $m^\mu_0$ is given by
\begin{eqnarray} \label{overlap_hetero_theory value}
  m_0^\mu = \mathrm{erf}\left(\frac{\gamma m_*^\mu}{\sqrt{2}\sigma_*}\right),
\end{eqnarray}
where $\mathrm{erf}(x)$ is the error function. See Appendix \ref{sec:HMM} for the derivation of (\ref{overlap_hetero_theory value}).

\subsection{Auto-associative memory model}
The AMM stores multiple patterns and can recall the most similar pattern when a pattern is input to the network. If the AMM stores $P$ patterns $\bm{\xi}^\mu=\left(\xi^\mu_1,\xi^\mu_2,\cdots,\xi^\mu_N\right)^\top, \mu=1,2,\cdots,P$, the synaptic weight $W^a_{ij}$ is given by
\begin{eqnarray} \label{W^a}
  W_{ij}^a &=& \frac{1}{N} \sum_{\mu=1}^P \xi_i^\mu \xi_j^\mu.
\end{eqnarray}
The state $x^{t+1}_i$ of the $i$-th neuron at time $t+1$ is given by
\begin{eqnarray} \label{auto_output}
  x^{t+1}_i &=& \mathrm{sgn}\left(h_i^t\right),
\end{eqnarray}
where $h^t_i$ is the internal state of the neuron at time $t$, given by
\begin{eqnarray} \label{internal state_auto}
  h_i^t &=& \sum_{j\neq i}^N W^a_{ij} x^t_j.
\end{eqnarray}

The overlap $m^\mu_t$ between the state $\bm{x}^t$ of the neurons in the AMM at time $t=0,1,2,\cdots$ and the $\mu$-th pattern $\bm{\xi}^{\mu}$ is defined as,
\begin{eqnarray} \label{overlap_auto}
  m^\mu_t &=& \frac{1}{N} \sum_{i=1}^N \xi_i^\mu x_i^t.
\end{eqnarray}
The overlap at time $t=0$ is the same as (\ref{overlap_hetero}) and is called the initial overlap. Now, assuming that the $\mu$-th pattern is recalled at time $t$, from equation (\ref{overlap_auto}), the internal state (\ref{internal state_auto}) can be expressed as,
\begin{eqnarray} \label{interference term_auto}
  h_i^t  &=& m^\mu_t \xi^\mu_i + z^t_i,
\end{eqnarray}
where the crosstalk noise term $z^t_i$ is given by
\begin{eqnarray} \label{crosstalknoise_auto}
  z^t_i &=& \frac{1}{N} \sum_{j\neq i}^N \sum_{\nu \neq \mu}^{P} \xi^\nu_i \xi^\nu_j x^t_j.
\end{eqnarray}
In the large system limit, the crosstalk noise term $z_i^t$ follows a Gaussian distribution with mean $0$ and variance $\sigma^2_t$, that is,
\begin{eqnarray}
  \sigma^2_t  &=& E\left[\left(z_i^t\right)^2\right].
\end{eqnarray}
The overlap $m_{t+1}^\mu$ and the variance $\sigma_{t+1}^2$ are given by
\begin{eqnarray} \label{overlap_auto_theory value}
  m_{t+1}^\mu &=& \text{erf}\left(\frac{m_t^\mu}{\sqrt{2}\sigma_t}\right),\\
  \label{sigma_auto_theory value}
  \sigma_{t+1}^2 &=& \alpha + U_{t+1}^2 \sigma_t^2 + 2\alpha\sum_{\tau=t-n+1}^t q_{t+1,\tau} \prod_{k=\tau+1}^{t+1} U_k . \nonumber \\
\end{eqnarray}
Because AMM is a recurrent neural network, the variance of the crosstalk noise has temporal correlations.~\cite{okada1995hierarchy,kawamura1999dynamics} See Appendix \ref{sec:AMM} for the derivation of (\ref{overlap_auto_theory value}) and (\ref{sigma_auto_theory value}).

\section{Associative watermarking method}
\begin{figure}[tb] \centering
  \includegraphics[width=80mm]{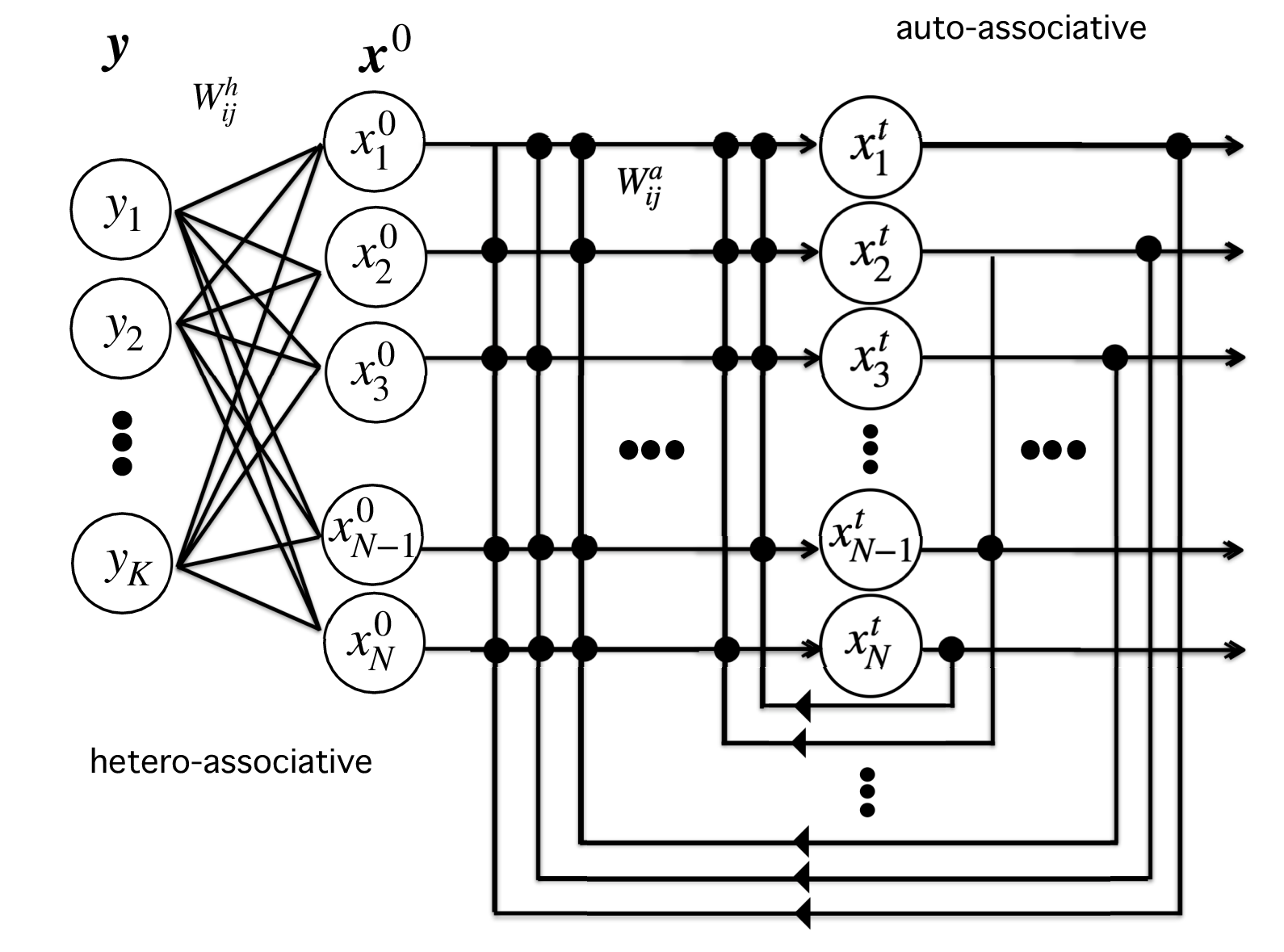}
  \caption{Structure of the associative watermarking memory} \label{fig:model}
\end{figure}

The AWM~\cite{kanegae2022proposal} introduces HMM for mapping features to watermarks in the zero-watermarking method and AMM for watermark error correction. When mapping multiple images and watermarks, the $\mu$-th feature $\bm{\eta}^{\mu}$ and watermark $\bm{\xi}^{\mu}$ in the zero-watermarking method correspond to the key pattern $\bm{\eta}^{\mu}$ and the associative pattern $\bm{\xi}^{\mu}$ in the HMM, respectively. Also, all secret keys $\bm{W}^{\mu}$ are represented by a single synaptic weight $\bm{W}^h$. Therefore, the AWM has several advantages compared to the zero-watermarking method. First, it does not need to manage the mapping between many secret keys and features—only the synaptic weight. Second, errors in the watermark can be corrected even if a feature is degraded. Third, the bit length of the feature $\bm{\eta}$ and that of the watermark $\bm{\xi}$ may be different. Fourth, the AWM has the same structure as the HASP~\cite{hirai1983model,kawamura1999dynamics}, as shown in Figure~\ref{fig:model}, and the retrieval process and memory capacity can be analyzed by using statistical neurodynamics~\cite{kawamura1999dynamics}.

\subsection{Proposed model} 
Let $t=-1$ be the time when the input $\bm{y}$ is given to the HMM, and let $t=0$ be the time when the output $\bm{x}^0$ of the HMM is input to the AMM.

Suppose that there are $P$ images and that the $N$-bit watermark $\bm{\xi}^{\mu}=\left(\xi^{\mu}_1,\xi^{\mu}_2,\cdots,\xi^{\mu}_N\right)^{\top}$ is associated with the $\mu$-th image. Similar to the zero-watermarking method~\cite{dong2011robust}, $K$ low-frequency components are extracted from the DCT coefficients of the $\mu$-th image, excluding the DC components. The extracted DCT coefficients are binarized by (\ref{sgn}) to obtain the features $\bm{\eta}^{\mu}=\left(\eta_1^{\mu},\eta_2^{\mu},\cdots,\eta_K^{\mu}\right)^{\top}$.

The features $\bm{\eta}^{\mu}$ and watermarks $\bm{\xi}^{\mu}$ are regarded as key patterns and associative patterns, respectively. The synaptic weights $\bm{W}^h$ and $\bm{W}^a$ are given by (\ref{W^h}) and (\ref{W^a}), respectively. The mapping between features and watermarks can be managed by storing the weight $\bm{W}^h$, and the watermarks can be recalled almost without error by storing the weights $\bm{W}^a$.

Suppose that the $\mu$-th image is attacked and that a feature is degraded. In the HMM, the sum of the products of the degraded feature $\tilde{\bm{\eta}}^{\mu}=\left(\tilde{\eta}^{\mu}_1,\tilde{\eta}^{\mu}_2,\cdots,\tilde{\eta}^{\mu}_K\right)^{\top}$ and the synaptic weight $\bm{W}^h$ is calculated by (\ref{internal state_hetero}) to obtain the output $\bm{x}^0$. This output $\bm{x}^0$ is given as the initial value of the AMM. The state of the neurons in the AMM at each time is determined by (\ref{auto_output}). The watermark $\bm{\xi}^{\mu}$ is known to be successfully recalled if the overlap between the state of the neurons $\bm{x}^t$ and the watermark $\bm{\xi}^{\mu}$ at time $t$ is $m^{\mu}_t=1$ after sufficient time.

\subsection{The Recall Process and Basin of Attraction}
The AWM corresponds to the one-to-one association case of the HASP~\cite{kawamura1999dynamics}. The recall process and memory capacity of this model were analyzed~\cite{kawamura1997storage,kawamura1999dynamics}. We evaluated the performance of our method using the macroscopic state equations derived in \ref{sec:HMM} and \ref{sec:AMM} and using computer simulations.

\begin{figure*}[tb] \centering
  \begin{minipage}{0.48\textwidth}\centering
    \includegraphics[width=80mm]{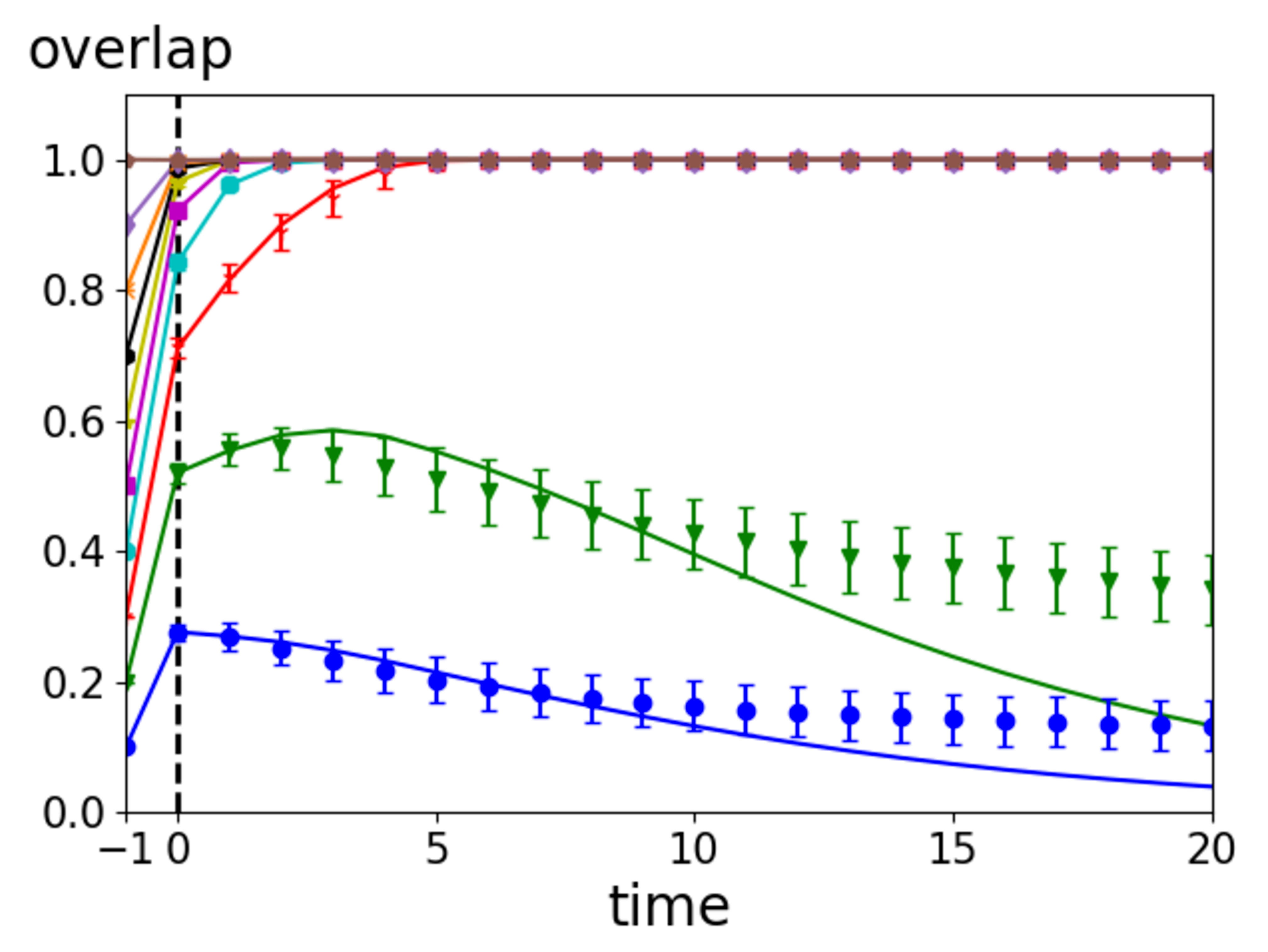} \\ 
    (a) loading rate $\alpha=0.08$ 
  \end{minipage}\hfill
  \begin{minipage}{0.48\textwidth}\centering
    \includegraphics[width=80mm]{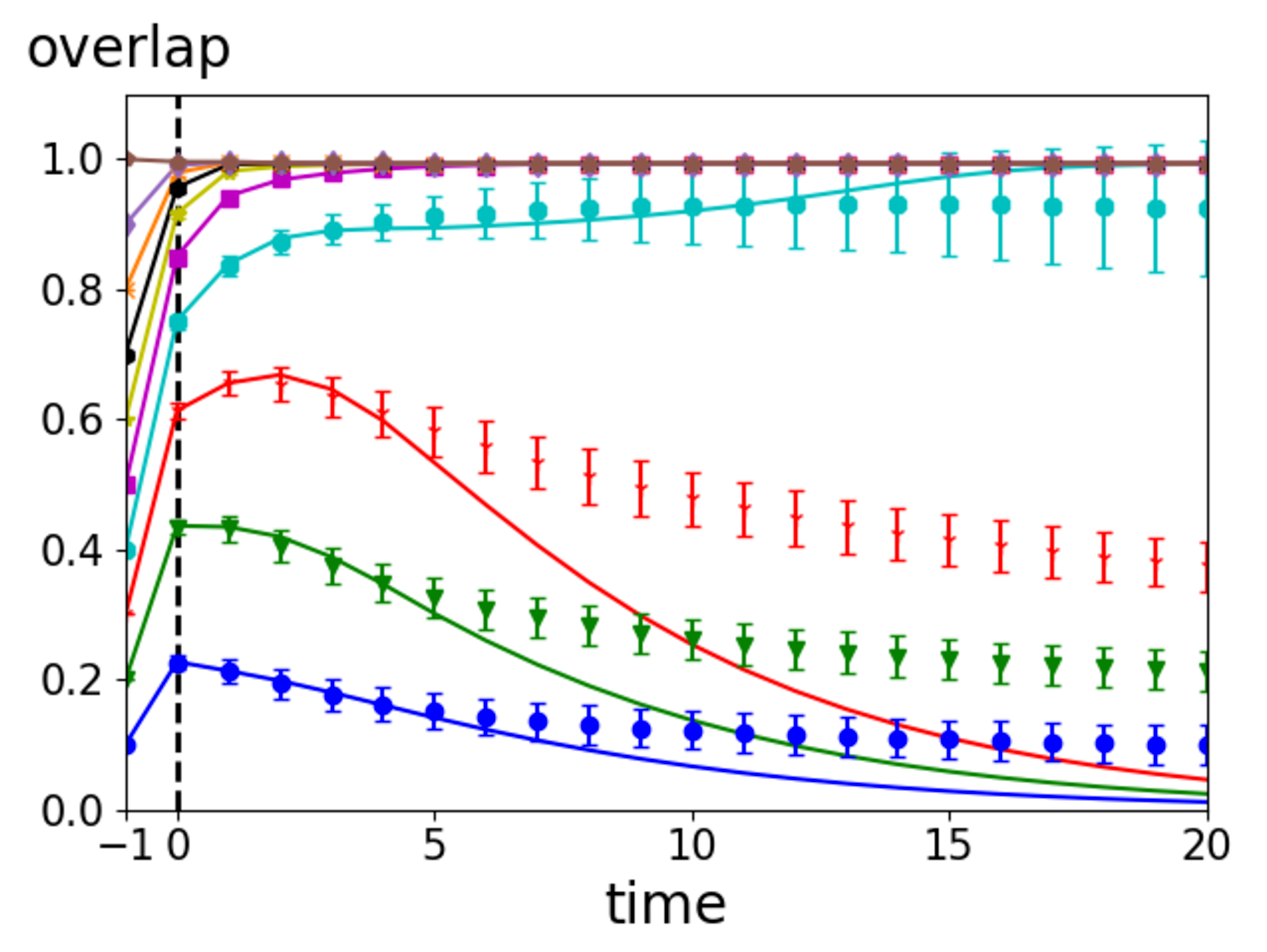} \\
    (b) loading rate $\alpha=0.12$ 
  \end{minipage}
  \caption{(Color online) Time evolution of overlaps $m_*^\mu, m_t^\mu$ for the loading rates $\alpha=0.08,0.12$ and $\gamma=1.0$.}
  \label{fig:overlap}
\end{figure*}

First, we evaluated the recall process of watermarks from degraded features. Figure~\ref{fig:overlap} shows the watermark recall process. The horizontal axis represents time $t$, and the vertical axis represents the overlap. Here, the overlap at time $t=-1$ represents the overlap $m_*^{\mu}$ for the feature $\bm{\eta}^{\mu}$, and the overlap at $t=0,1,2,\cdots$ represents the overlap $m_t^{\mu}$ for the watermark $\bm{\xi}^{\mu}$. The solid line represents the theoretical value of overlap represented by (\ref{overlap_hetero_theory value}) and (\ref{overlap_auto_theory value}). The cases recalled from various initial overlaps are shown. In the computer simulation, the features and watermarks were randomly generated by (\ref{eq:Peta}) and (\ref{eq:Pxi}). The degraded figures input to the AWM were flipped to be the given initial overlap. The bit lengths of the features and watermark were set to $N=10000, K=10000$, respectively. That is, $\gamma=1.0$. The overlap at each time was illustrated by an error bar. The error bars represent the mean and the standard deviation for $100$ trials. (a) and (b) show the results when the number of patterns was $P=800\; (\alpha=0.08)$ and $P=1200\; (\alpha=0.12)$, respectively.

Figure~\ref{fig:overlap} shows the overlap $m_0^{\mu}$ of the output layer of the HMM was larger than the overlap $m_*^{\mu}$ of the input layer. This means that the HMM can not only store the mapping between features and watermarks but also correct watermark errors. The recallable boundary value is called the critical overlap $m_c$. For $\alpha=0.08$ in Figure~\ref{fig:overlap} (a), the critical overlap was $m_c>0.2$ for the fourth-order theory ($n=4$). Computer simulations also showed that $m_c>0.2$. Similarly, for $\alpha=0.12$ in Figure~\ref{fig:overlap} (b), both theory and computer simulations showed $m_c>0.3$. These results revealed that fourth-order theory can quantitatively represent the results of computer simulations.

\begin{figure*}[tb]  \centering
    \begin{minipage}{0.48\textwidth}
        \centering
        \includegraphics[width=80mm]{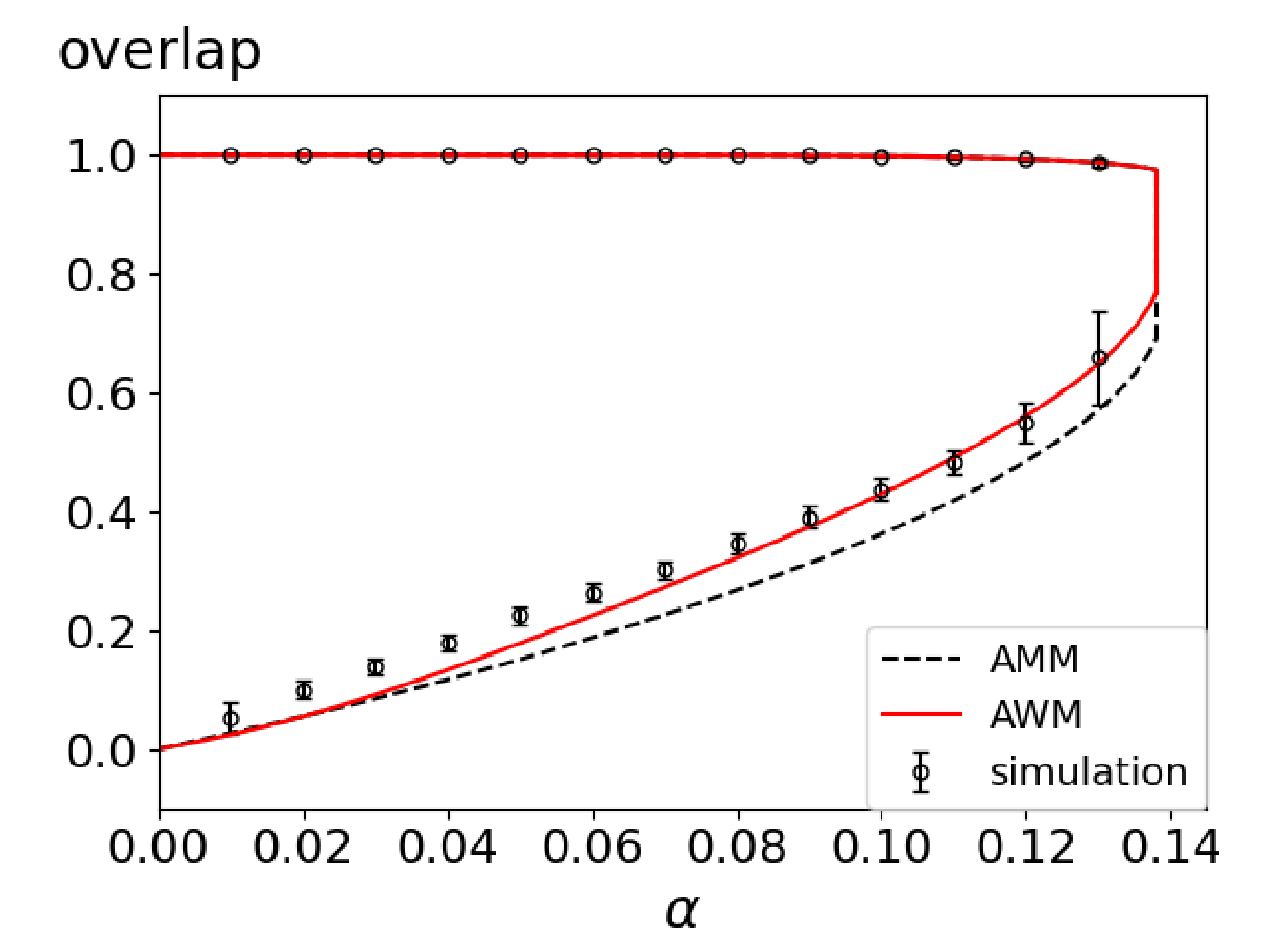}
        (a)\(\ K=5000,N=10000\ (\gamma=0.5)\) 
    \end{minipage}
    \hfill
    \begin{minipage}{0.48\textwidth}
        \centering
        \includegraphics[width=80mm]{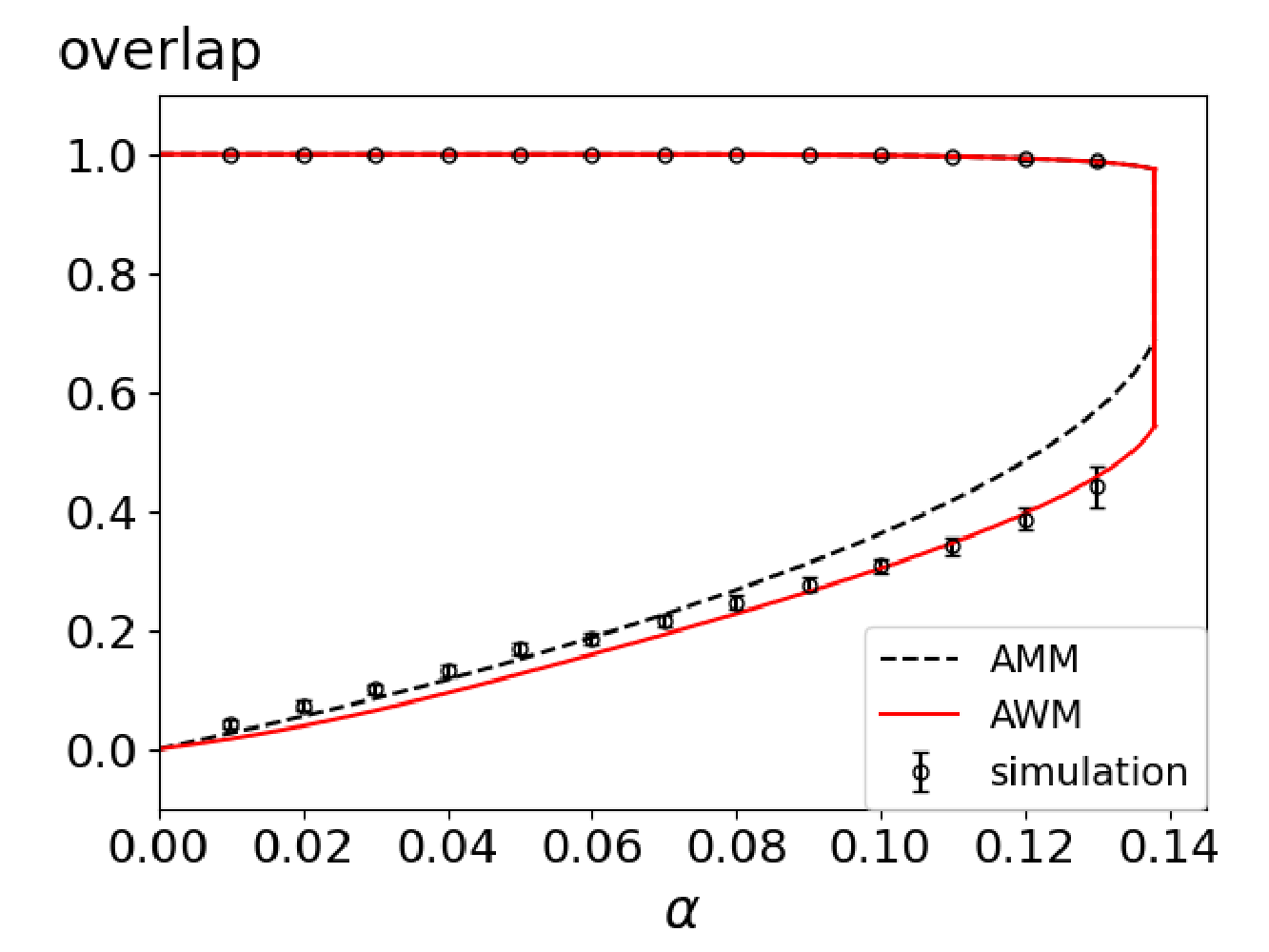}
        (b)\(\ K=N=10000\ (\gamma=1.0)\)
    \end{minipage}
    \caption{(Color online) Basin of attraction for the AMM and AWM}
    \label{fig:basin of attraction}
\end{figure*}

Figure~\ref{fig:basin of attraction} shows the basins of attraction to evaluate the error correction capability of AWM. (a) is a case where the bit lengths of the features and watermark were $K=5000, N=10000,$ i.e., $\gamma=0.5$, respectively, and (b) is a case where $K=N=10000\,(\gamma=1.0)$. The horizontal axis represents the loading rate $\alpha$, and the vertical axis represents the overlap. The lower curves show the critical overlaps $m_c$. The upper curves show the equilibrium overlaps $m_{\infty}$ when the network was in equilibrium at time $t\to\infty$. The equilibrium overlap represents the stability of the memorized pattern and is the overlap at time $t\to\infty$ when the network is recalled from an initial state with initial overlap $m_0=1$. The region bounded by the equilibrium overlap and the critical overlap is the basin of attraction.

We can see from Fig.~\ref{fig:overlap} that the overlaps were well converged by $t=20$. Therefore, the equilibrium overlaps obtained from the computer simulations were the overlap $m_{20}$ at $t=20$. The error bars were the results obtained from the computer simulations, and they represent the mean and standard deviation of the overlap $m_{20}^{\mu}$ over $100$ trials at each loading rate $\alpha$. The black and red dashed lines are the results obtained from the fourth-order theory for AMM and AWM, respectively. The results of the computer simulations were in good agreement with the theoretical curves of the AWM. In Fig.~\ref{fig:basin of attraction} (a), the basin of attraction of AWM was lower than that of AMM, while the basin of attraction of AWM was higher in (b). As the bit length ratio of the features and watermarks, $\gamma=K/N$, decreases, the overlap of the HMM's output decreases from (\ref{overlap_hetero_theory value}). In other words, when $\gamma$ is low, the error correction capability of the HMM is low.

\subsection{Amount of information required to store}
Let us calculate the amount of information required to store the secret keys, $C$, in AWM. Information on the weight matrices $W^h_{ik}$ and $W^a_{ij}$ must be stored. If the activation function in (\ref{hetero_output}) and (\ref{auto_output}) is $\mathrm{sgn}()$, the division by $N$ operation in (\ref{W^h}) and (\ref{W^a}) can be omitted. Thus, the weights can be stored as integers. The maximum value of $W_{ik}^h$ and $W_{ij}^a$ is $P$ if the values of the patterns are $\eta_k^\mu=1$ and $\xi_i^\mu=1$ for all $\mu$. In practice, the maximum is less than $P$ because this case is unlikely. It takes $\log_2P$ bits to store the value of a weight. Since $W_{ik}^h$ is an asymmetric matrix, $C_h=NK\log_2P$ bits are required to store this weight. In addition, since $W_{ij}^a$ is a symmetric matrix and the diagonal components are not used, $C_a = (N(N-1)/2)\log_2P$ bits are required. Accordingly, the total information to be stored is $C=C_h+C_a=N(K+(N-1)/2)\log_2P$ bits. Let us compare the zero-watermarking method and AWM in terms of the amount of information. Obviously, the zero-watermarking method requires less information than AWM. Note, however, that the proposed AWM does not require indexing. In other words, the proposed method is suitable for managing multiple images.

\section{Performance Evaluation of the AWM}
The order parameter equations are equivalent to those of the HASP~\cite{kawamura1999dynamics}. The previous section presented an evaluation of the performance of AWM with the basin of attraction as an AMM. This section features an evaluation of its performance as a watermarking model. It was evaluated quantitatively in terms of the bit error rate (BER) of the watermark. The BERs obtained from the zero-watermarking method, HMM, and AWM were evaluated using features obtained from images. The BER is defined by
\begin{equation}
  \text{BER}\left(m\right) = \frac{1-m}{2},
\end{equation}
where \(m\) is an overlap. Therefore, the BER for the \(\mu\)-th watermark at time \(t\) is represented by BER\(\left(m_t^\mu \right)\).

\subsection{JPEG compression attack}

We evaluated the BERs of the watermarks when the images were JPEG compressed and when the degraded features were extracted. JPEG compression was applied to 1200 original images~\cite{tarunpathak_2021} of $512\times512$ pixels. The bit lengths of the features were set to $K=5000, 10000$. The watermarks were randomly generated by (\ref{eq:Pxi}). The bit length of the watermarks was set to $N=10000$. That is, $\gamma=0.5,1.0$ and $\alpha=0.12$. With weak JPEG compression, the watermark could be recalled correctly due to the error correction capability of the HMM. Therefore, we examined the case of strong compression, where the watermark was incorrectly recalled at the output of the HMM. Here, the quantization level of JPEG compression was set to $Q=5$.

\begin{figure*}[tb]  \centering
    \begin{minipage}{0.48\textwidth}
        \centering
        \includegraphics[width=80mm]{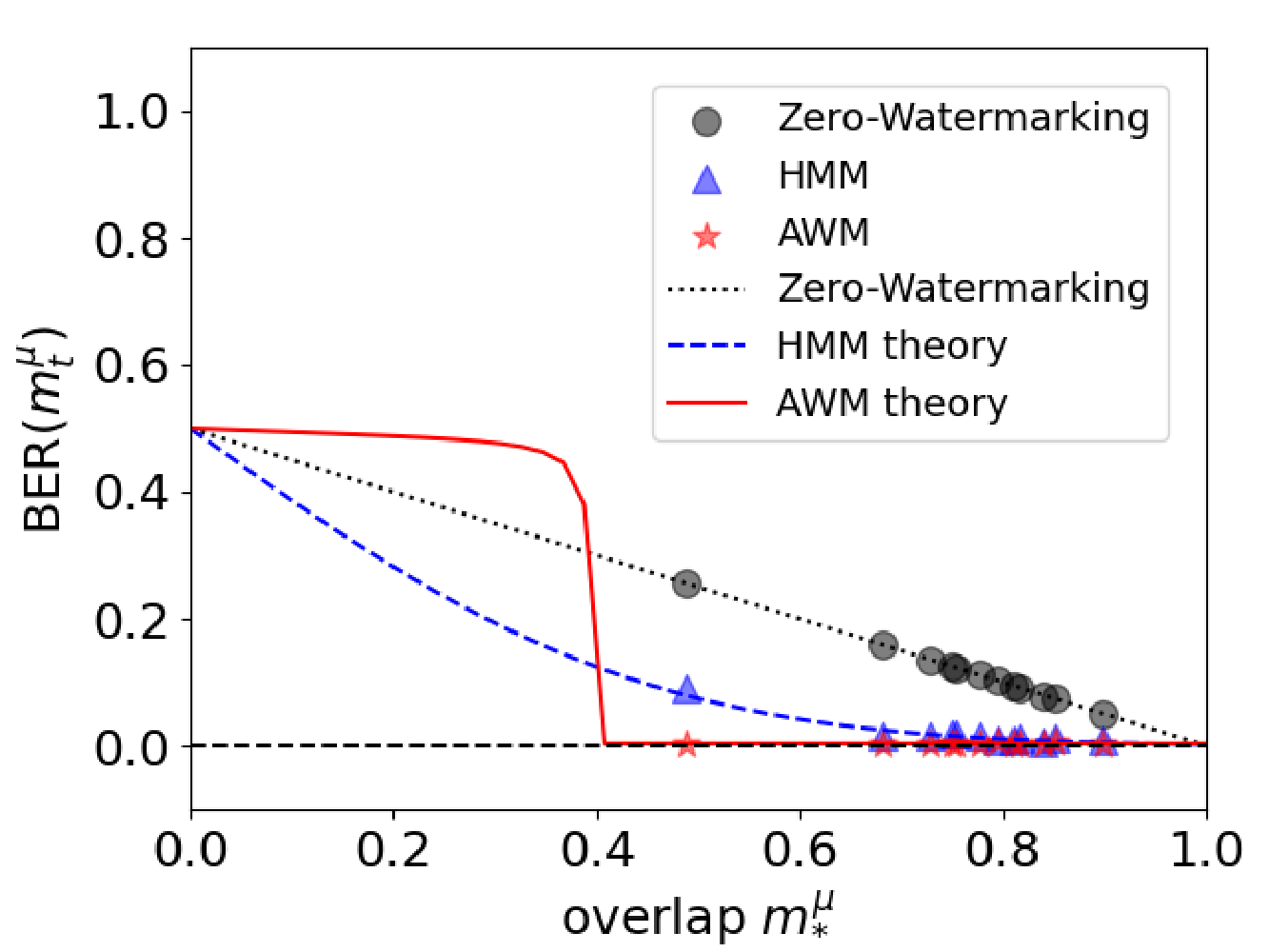}
        (a)\(\ K=5000,N=10000\ (\gamma=0.5)\) 
    \end{minipage}
    \hfill
    \begin{minipage}{0.48\textwidth}
        \centering
        \includegraphics[width=80mm]{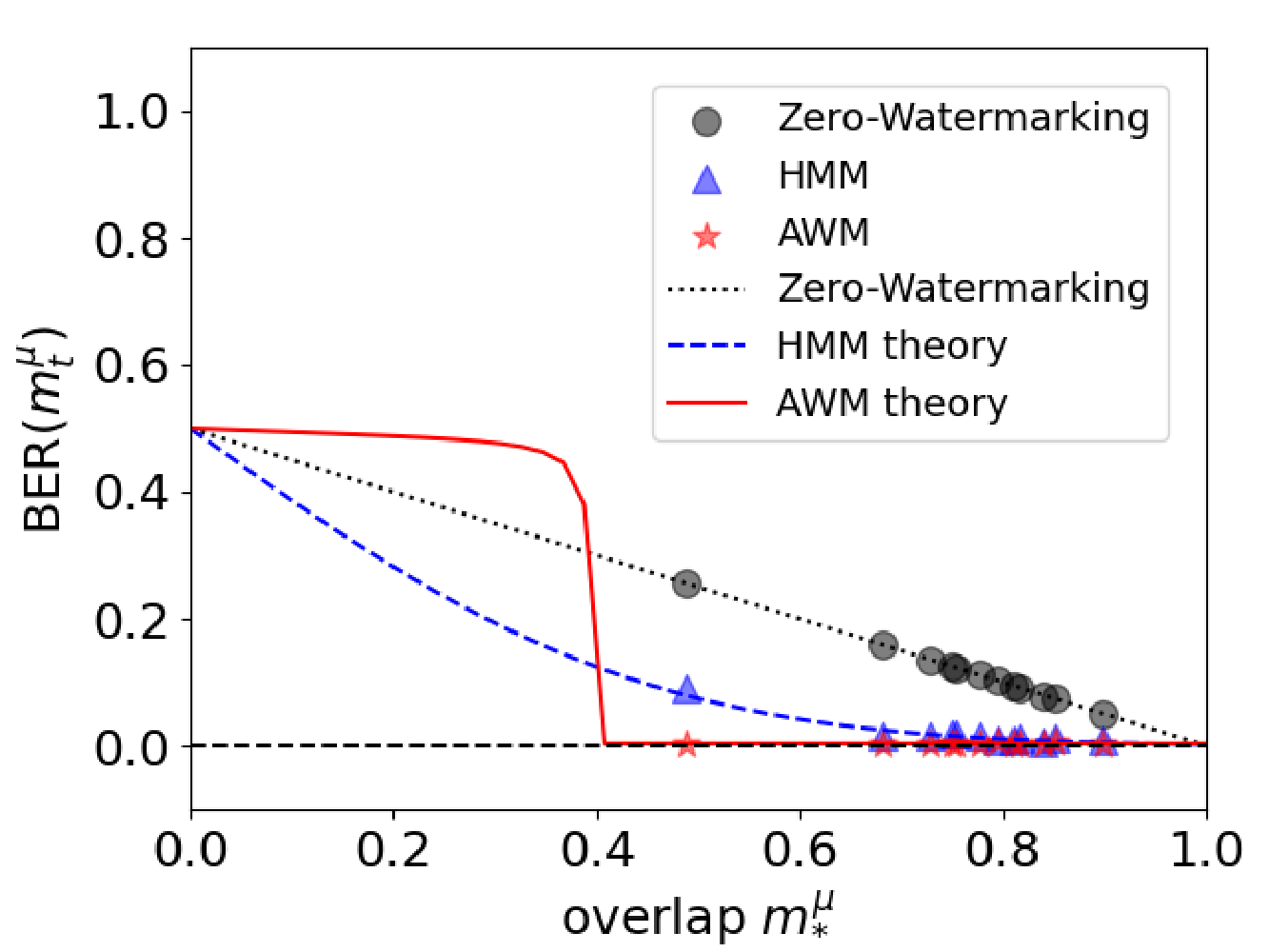}
        (b)\(\ K=N=10000\ (\gamma=1.0)\)
    \end{minipage}

    \caption{(Color online) BER of the zero-watermarking method, HMM, and AWM for JPEG compression attack \(\left(\alpha=0.12\right)\)}
    \label{fig:ber vs overlap}
\end{figure*}

The BERs of the watermarks are shown in Fig.~\ref{fig:ber vs overlap} when the images were JPEG compressed. The horizontal and vertical axes represent the overlap $m_*^{\mu}$ of the feature $\bm{\eta}^{\mu}$ and the $\mathrm{BER}\left(m_t^{\mu} \right)$ of the watermark $\bm{\xi}^{\mu}$, respectively. Because the degraded feature $\tilde{\bm{\eta}}^{\mu}$ was given as input $\bm{y}$ of the HMM, the overlap $m_*^{\mu}$ was calculated using the (\ref{overlap_feature}). The blue and red solid lines represent, respectively, the theoretical BER for the output of the HMM obtained by the (\ref{overlap_hetero_theory value}) and the output of the AWM at time $t=20$ obtained by the (\ref{overlap_auto_theory value}). The black, blue, and red points represent the BERs of the watermarks obtained from the zero-watermarking model, the HMM, and the AWM obtained from computer simulations. The overlaps for AWM were the value of $m_{20}^\mu$ at time $t=20$. The BERs were plotted using 12 images compressed at a high compression ratio ($Q=5$) so as to generate errors in the output of the HMM. Note that it is difficult to obtain a small overlap $m_*^\mu$ even with a high compression rate. Figure~\ref{fig:ber vs overlap} (a) shows the results for bit lengths of $K=5000, N=10000$ ($\gamma=0.5$) and (b) for those of $K=N=10000$ ($\gamma=1.0$), where the loading rate was $\alpha=0.12$.
The theoretical curves showed that the BER of the HMM varied gradually, while the BER of the AWM varied sharply at the critical overlap $m_c$, owing to the AMM. Because of their error correction capabilities, both the HMM and AWM had smaller errors than the zero-watermarking method. In addition, errors that could not be corrected by the HMM could be corrected by the AMM. The error was almost zero for the AWM in total. The overlap for the HMM appeared to be out of the theoretical curve because the features extracted from the images did not satisfy the condition (\ref{eq:Peta}).

\subsection{Gaussian noise attack}
\begin{figure}[tb] \centering
  \includegraphics[width=85mm]{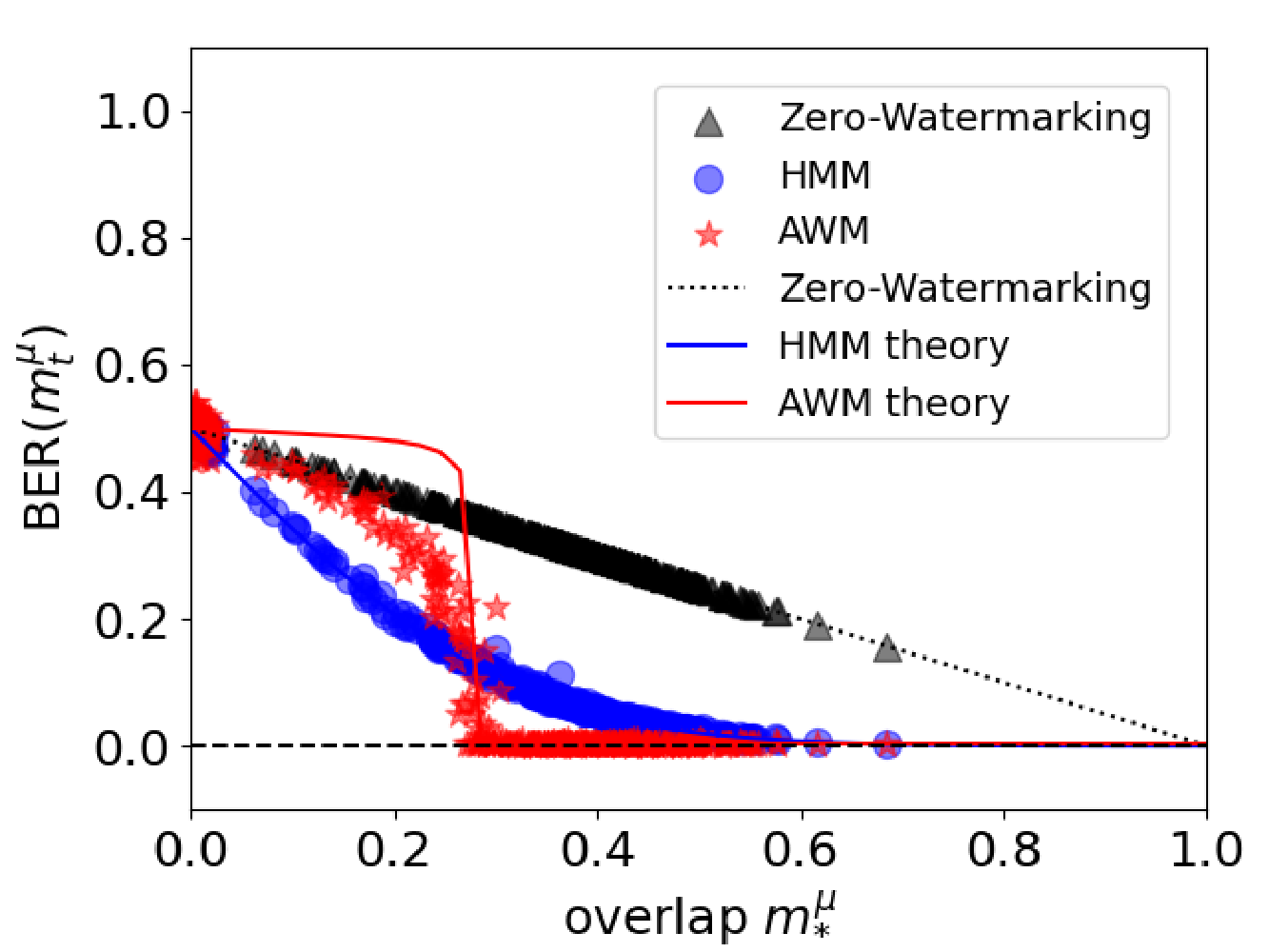}
  \caption{(Color online) BER for Gaussian noise attack ($\alpha=0.12, \gamma=2.0$)}
  \label{fig:ber_AWGN}
\end{figure}

Next, we show the results for additive Gaussian noise, since the JPEG compression attack did not degrade the watermarks below the critical overlap. Gaussian noise with a mean of 100 and a standard deviation of 100 was used to generate large errors. Although pixel values usually take 256 gray levels, we assume that the pixel values in this experiment take values beyond that range. Figure~\ref{fig:ber_AWGN} shows the BER in the case of a Gaussian noise attack, where $K=20000, N=10000$ $(\gamma=2.0)$, and $P=1200\;(\alpha=0.12)$. There are differences in BER for AWM below the critical overlap because theory and computer simulations do not match below the critical overlap, as shown in Fig~\ref{fig:overlap}. However, they are in good agreement above the critical overlap. If the image is highly degraded and the overlap of the watermark is below the critical overlap (which is rarely the case), only the HMM performs better because the BER of the HMM is lower than that of the AWM.

\section{Conclusion}
While the zero-watermarking method does not degrade the original image, it cannot correct watermark errors. Also, managing the mapping between features and secret keys becomes difficult because the number of secret keys increases as the number of images increases. However, the mapping between the features and secret keys in the associative watermarking model can be managed collectively by simply storing synaptic weights. In addition, the features and watermarks have no restrictions on the bit length. Furthermore, even if the image is attacked, the watermark can be corrected and extracted~\cite{kanegae2022proposal}. In this paper, we presented an evaluation of the BER in the AWM through theory and computer simulations. The macroscopic state equations for the AWM were equivalent to the one-to-one associative case of the HASP, and the retrieval process and storage capacity were calculated using statistical neurodynamics. The retrieval processes obtained from fourth-order theory and those from computer simulations were in good agreement. The storage capacity of the AWM was the same as that of the AMM. We also found that the basin of attraction for the AWM varied in size depending on the ratio $\gamma=K/N$ of the bit lengths of the features and the watermark.

Actual images were usually JPEG compressed, and the BER of the extracted watermark was evaluated to assess the effectiveness of AWM as a watermarking model. The results showed that the error rates of both the HMM and AWM were less than those of conventional zero-watermarking. Furthermore, the BERs of the AWM were almost zero. As a result, our method extends the capabilities of the zero-watermarking method and has excellent functions. Although we discussed the JPEG compression and additive Gaussian noise attacks, the results would be the same if other image processing were applied. The performance was determined by the overlap at the input to the HMM.

When processing multiple images, the independence of the features must be taken into account. For example, in the case of chest X-ray images, since they are all similar images, they are highly correlated with each other. The extraction of independent features from such correlated images is an open problem.

\begin{acknowledgment}
This work was supported by JSPS KAKENHI Grant Number 20K11973 and 24K15106.
The computations were performed using the supercomputer facilities at the
Research Institute for Information Technology, Kyushu University.
\end{acknowledgment}

\appendix
\section{Order Parameter Equations}
To analyze the dynamics of the recall process in the associative memory model, we derived a macroscopic state equation for this model using statistical neurodynamics~\cite{amari1988statistical,okada1995hierarchy}. This theory describes the state of the network using multiple macroscopic state variables, and the recall process is represented by the macroscopic state equations. Okada's theory~\cite{okada1995hierarchy} considers temporal correlations of states up to $n$ previous time. In this paper, we call it $n$-order theory. For $n=1$, it is equivalent to the Amari-Maginu theory~\cite{amari1988statistical}. The storage capacity converges to the results obtained from the equilibrium state theory~\cite{amit1985storing} as the order $n$ increases.

\subsection{Macroscopic state equations for the HMM} \label{sec:HMM}
We herein discuss the case where the $\mu$-th pattern is recalled. The superscript for $\mu$ is omitted.
macroscopic state equations are derived for the HMM. From (\ref{crosstalknoise_hetero}), the mean of the crosstalk noise term at time $t=-1$ is 0, and its variance is given by
\begin{eqnarray}
  \sigma^2_* &=& E\left[\left(z_i^*\right)^2\right] = \alpha \gamma,
\end{eqnarray}
where $\alpha=P/N$ and where $\gamma=K/N$. The macroscopic state equations at time $t=0$ are given by
\begin{eqnarray}
  \label{overlap_hetero_equation}
  m_0 &=& \int D_z \left\langle\xi\mathrm{sgn}\left(\gamma\xi m_* +\sigma_* z\right) \right\rangle_{\xi},\\
  \sigma_0^2 &=& \alpha + \sigma_*^2 U_0^2,\\
  \label{U_hetero_equation}
  U_0 &=& \frac{1}{\sigma_*} \int D_z\ z \left\langle \mathrm{sgn}\left(\gamma\xi m_* +\sigma_*z\right) \right\rangle_\xi,
\end{eqnarray}
where $\left\langle\cdot\right\rangle_{\xi}$ represents the average over $\mu$-th pattern $\xi^{\mu}$ and where $D_z$ is defined by
\begin{eqnarray}
  D_z = \frac{dz}{\sqrt{2\pi}} \exp\left( -\frac{z^2}{2} \right).
\end{eqnarray}
These equations can be solved as
\begin{eqnarray}
  m_{0} &=& \mathrm{erf}\left(\frac{\gamma m_*}{\sqrt{2}\sigma_*}\right),\\
  U_{0} &=& \sqrt{\frac{2}{\pi\sigma_*^2}}\exp\left(-\frac{\gamma^2 m_*^2}{2\sigma_*^2} \right).
\end{eqnarray}

\subsection{Macroscopic state equations for the AMM} \label{sec:AMM}
The macroscopic state equations for the AMM were derived by Okada~\cite{okada1995hierarchy}. The $n$-order macroscopic state equations at time $t\geq0$ are given as follows.
\begin{eqnarray}
  m_{t+1} &=& \int D_z \left\langle \xi\mathrm{sgn}\left(\xi m_t +\sigma_t z\right) \right\rangle_\xi ,\label{overlap_auto_equation} \\
  \sigma_{t+1}^2 &=& \alpha + U_{t+1}^2 \sigma_t^2 + 2\alpha\sum_{\tau=t-n+1}^t q_{t+1,\tau} \prod_{k=\tau+1}^{t+1} U_k ,   \label{sigma_auto_equation} \\
  U_{t+1} &=& \frac{1}{\sigma_t} \int D_z\ z\left\langle \mathrm{sgn}\left(\xi m_t +\sigma_t z \right) \right\rangle_\xi, \label{U_auto_equation}\\
  q_{t+1} &=& \int D_z \left\langle \mathrm{sgn}\left(\xi m_t +\sigma_t z\right)^2 \right\rangle_\xi. \label{q_auto_equation}
\end{eqnarray}
These equations can be solved as
\begin{eqnarray}
  m_{t+1} &=& \mathrm{erf}\left(\frac{m_t}{\sqrt{2}\sigma_t}\right),\\
  U_{t+1} &=& \sqrt{\frac{2}{\pi \sigma_t^2}}\exp\left( -\frac{m_t^2}{2\sigma_t^2} \right),\\
  q_{t+1} &=& 1.
\end{eqnarray}
The $q_{t+1,\tau}$ represents the correlation between the state $x_{t+1}$ at time $t+1$ and the state $x_{\tau}$ at time $\tau$, and it is defined as
\begin{eqnarray}
  q_{t+1,\tau} &=& E\left[x_i^{t+1}x_i^{\tau}\right].
\end{eqnarray}
This variable can be expressed as
\begin{align}
  q_{t+1,\tau} &= \int D_c\int D_a \int D_b
  \left\langle\mathrm{sgn}\left(\xi m_t +\sigma_t\left(d_0 a+d_1c\right)\right)\right. \nonumber\\
  &\times\left.\mathrm{sgn}\left(\xi m_{\tau-1} +\sigma_{\tau-1}\left(d_0b+d_1c\right) \right)
  \right\rangle_\xi,
\end{align}
where
\begin{equation}
  d_0 = \sqrt{1-\frac{C_{t,\tau-1}}{\sigma_t \sigma_{\tau-1}}}, \quad d_1 = \sqrt{\frac{C_{t,\tau-1}}{\sigma_t \sigma_{\tau-1}}},
\end{equation}
and where $C_{t, \tau-1}$ denotes the correlation $C_{t, \tau-1}=E[z_{t} z_{\tau-1}]$ between the crosstalk noise term $z_{t}$ at time $t$ and the term $z_{\tau-1}$ at time $\tau-1$. These correlations can be expanded as
\begin{align}
  C_{t,\tau-1} &=0 \;(\tau=t-n+1,n\geq1),\\
  C_{t,\tau-1} &=\alpha q_{t,\tau-1}+U_t C_{t-1,\tau-1} \;(\tau=t-n+2,n\geq2),\\
  C_{t,\tau-1} &=\alpha q_{t,\tau-1}+U_t U_{\tau-1}C_{t-1,\tau-2} \nonumber\\
  &+\alpha\sum_{\eta=\tau-n+1}^{\tau-2} q_{t,\eta}\prod_{k=\eta+1}^{\tau-1}U_k +\alpha\sum_{\eta=\tau-n+1}^{t-1} q_{\eta,\tau-1}\prod_{k=\eta+1}^{t}U_k \nonumber\\
  &(t-n+3\leq\tau\leq t,n\geq3).
\end{align}

\subsection{Temporal correlation between the HMM and the AMM}
The AWM consists of both the HMM and the AMM. Therefore, we need to consider the temporal correlation between the input $(t=-1)$ of the HMM and the state of the AMM at time $t$. However, the input at time $t=-1$ is the feature $\bm{\eta}^{\mu}$ and the state at time $t$ is the state recalling the watermark $\bm{\xi}^{\mu}$. Because the feature and the watermark are uncorrelated, the temporal correlation between the input and the state can be ignored. Therefore, the correlation of states at time $t=-1$ can be $q_{t+1,-1}=0$.

\end{document}